\documentclass[smallextended]{svjour3}       
\usepackage{amsmath}
\usepackage{amssymb}

\usepackage{graphicx}

\begin{document}

\title{Discrete phase space and continuous time relativistic quantum mechanics II: Peano circles, hyper-tori phase cells, and fibre bundles}

\author{Anadijiban Das         \and
        Rupak Chatterjee     
}

\institute{Anadijiban Das \at
              Department of Mathematics, Simon Fraser University, Burnaby, British Columbia, V5A 1S6, Canada \\
              \email{Das@sfu.ca}           
           \and
           Rupak Chatterjee \at
              Center for Quantum Science and Engineering\\
Department of Physics, Stevens Institute of Technology, Hoboken, NJ 07030, USA \\ \email{Rupak.Chatterjee@Stevens.edu}   
}

\date{Received: date / Accepted: date}

\maketitle

\begin{abstract}
The discrete phase space and continuous time representation of relativistic quantum mechanics is further investigated here as a continuation of paper \textbf{I} \cite{DasRCI}. The main mathematical construct used here will be that of an area-filling Peano curve. We show that the limit of a sequence of a class of Peano curves is a Peano circle denoted as $\bar{S}^{1}_{n}$, a circle of radius $\sqrt{2n+1}$ where $n \in \{0,1,\cdots\}$. We interpret this two-dimensional Peano circle in our framework as a phase cell inside our two-dimensional discrete phase plane. We postulate that a first quantized Planck oscillator, being very light, and small beyond current experimental detection, occupies this phase cell $\bar{S}^{1}_{n}$.  The time evolution of this Peano circle sweeps out a two-dimensional vertical cylinder analogous to the world-sheet of string theory. Extending this to three dimensional space, we introduce a $(2+2+2)$-dimensional phase space hyper-tori $\bar{S}^{1}_{n^1} \times \bar{S}^{1}_{n^2} \times \bar{S}^{1}_{n^3}$ as the appropriate phase cell in the physical dimensional discrete phase space. A geometric interpretation of this structure in state space is given in terms of product fibre bundles.

We also study free scalar Bosons in the background $[(2+2+2)+1]$-dimensional discrete phase space and continuous time state space using the relativistic partial difference-differential Klein-Gordon equation. The second quantized field quantas of this system can cohabit with the tiny Planck oscillators inside the $\bar{S}^{1}_{n^1} \times \bar{S}^{1}_{n^2} \times \bar{S}^{1}_{n^3}$ phase cells for eternity. Finally, a generalized free second quantized Klein-Gordon equation in a higher $[(2+2+2)N+1]$-dimensional discrete state space is explored. The resulting discrete phase space dimension is compared to the significant spatial dimensions of some of the popular models of string theory.
 

\end{abstract}

\section{Introduction}

We begin in section 2 by introducing the concept of a Peano curve \cite{Clark,Gelbaum}. In 1890, G. Peano startled the mathematical world with the introduction of an area-filling curve. After that, many area-filling curves have been discovered including those by David Hilbert. To our knowledge, very few papers in the mathematical physics arena have used this concept \cite{DasHaldar}.

We begin by discussing the original Peano curve that fills a square $\bar{D}$ of unit area inside $\mathbb{R}^2$. In section 3, we derive other Peano curves filling a sequence $\{\bar{D}_M\}^{\infty}_{M=1}$ of unit areas each in the shape of a rectangle. Next, we introduce a double sequence $\{\bar{D}_{Mn}\}^{\infty}_{M=1}$ for $n \in \{0,1,2,...\}$. Each of closed regions $\bar{D}_{Mn}$ for a fixed $n$ is endowed with a unit area and a rectangular shape. Moreover, each of these regions $\bar{D}_{Mn}$ is covered by a Peano curve. Next, we discuss another  double sequence $\{\bar{A}_{Mn}\}$ of closed regions where each region $\bar{A}_{Mn}$ is annular in shape and possesses a unit area. This annular region $\bar{A}_{Mn}$ is also covered by a Peano curve. Finally, in the limiting process of $M \rightarrow \infty$ for a fixed $n$, the annular region $\bar{A}_{Mn}$  collapses into a circle $\bar{S}^{1}_{n}$ of unit area. Since this cirlce contains a 'squashed' Peano curve, one calls $\bar{S}^{1}_{n}$ a Peano circle. Section 4 details more analysis on this Peano circle. 

The simplified Klein-Gordon equation in the background of a $[(1+1)+1]$-dimensional discrete phase space and continuous time is discussed in section 5. Some physically relevant results are also derived.

In section 6, we introduce the abstract concept of fibre bundles \cite{Choquet} and discuss their applications to certain physical problems. We apply this concept in section 7 to the physically important $[(2+2+2)+1]$-dimensional phase space. This discrete phase space and continuous time arena has been investigated with respect to the second quantization of many free relativistic field theories \cite{DasI,DasII}. The corresponding formulations of interacting relativistic second quantized fields yielding an $S^{\#}$-matrix series were given in \cite{DasIII}. Second order expansion terms of this $S^{\#}$-matrix involving M\o ller scattering of quantum electrodynamics \cite{Jauch,Peskin} produced a non-singular Coulomb potential in \cite{DasBen,DasRC}.

In section 7, we define a new set-theoretic mapping \cite{Goldberg} by the direct or Cartesian product \cite{Lightstone} as:
\begin{equation}
\begin{array}{c}
\bar{S}^{1}_{n^1} \times \bar{S}^{1}_{n^2} \times \bar{S}^{1}_{n^3} = \displaystyle{\lim_{M \to \infty}} \\
\\
\left([g^{M}_{n^1} \times h^{M}_{n^1} \times h^{M}_{1}] \times [g^{M}_{n^2} \times h^{M}_{n^2} \times h^{M}_{2}] \times [g^{M}_{n^3} \times h^{M}_{n^3} \times h^{M}_{3}] \right) \left( \bar{D}_{1} \times \bar{D}_{2} \times \bar{D}_{3} \right) \\
\\
\subset \mathbb{R}^2  \times \mathbb{R}^2 \times \mathbb{R}^2.
\end{array}
\end{equation}
The left hand side of this equation ($\bar{S}^{1}_{n^1} \times \bar{S}^{1}_{n^2} \times \bar{S}^{1}_{n^3} $) is a hyper-torus \cite{Massey} of physical dimension $\left[\dfrac{ML^2}{T}\right]^3$. It will be shown that these hyper-tori can represent a phase cell in the usual $(2+2+2)$-dimensional discrete phase space. 

In section 8, the second quantization of the scalar field $\Phi(n^1, n^2, n^3;t) =: \Phi(\mathbf{n};t)$ is discussed in the arena of $[(2+2+2)+1]$-dimensional discrete phase space and continuous time using the relativistic Klein-Gordon operator equation. A partial difference-differential version of the Klein-Gordon equation is solved for a special class of solutions involving Fourier-Hermite transforms \cite{DasII}. The total energy $\mathcal{H}$, momentum component $\mathcal{P}_j$ and electric charge $\mathcal{Q}$ are worked out for the second quantized Klein-Gordon scalar field operator $\Phi(\mathbf{n};t)$. 

Finally, in section 9, we study discrete $(2+2+2)N$-dimensional phase space many-particle systems and the corresponding Peano hyper-tori $\bar{S}^{1}_{n^1} \times \bar{S}^{1}_{n^2} \times \cdots \bar{S}^{1}_{n^N}$. Each of these Peano hyper-tori has the physical dimension of $\left[\dfrac{ML^2}{T}\right]^N$ and represent appropriate phase cells inside the discrete  $(2+2+2)N$-dimensional phase space. The generalized Klein-Gordon equation within this discrete $[(2+2+2)N +1]$-dimensional phase space and continuous time arena is also discussed. As our $2N$-dimensional Peano hyper-tori is a phase-cell inside a discrete $6N$-dimensional phase space, we compare our structure with that of the spatial dimension of string theory \cite{Green,PolchinskiI,PolchinskiII}.

\section{The area filling curve of Peano}

Consider a continuous, piece-wise linear, oriented, parametrized curve \cite{Clark} $f^{1}(u)$ inside a $(1+1)$-dimensional plane $\mathbb{R}^2$ as depicted in figure 1. 
\begin{figure}
\begin{center}
\includegraphics[scale=0.16]{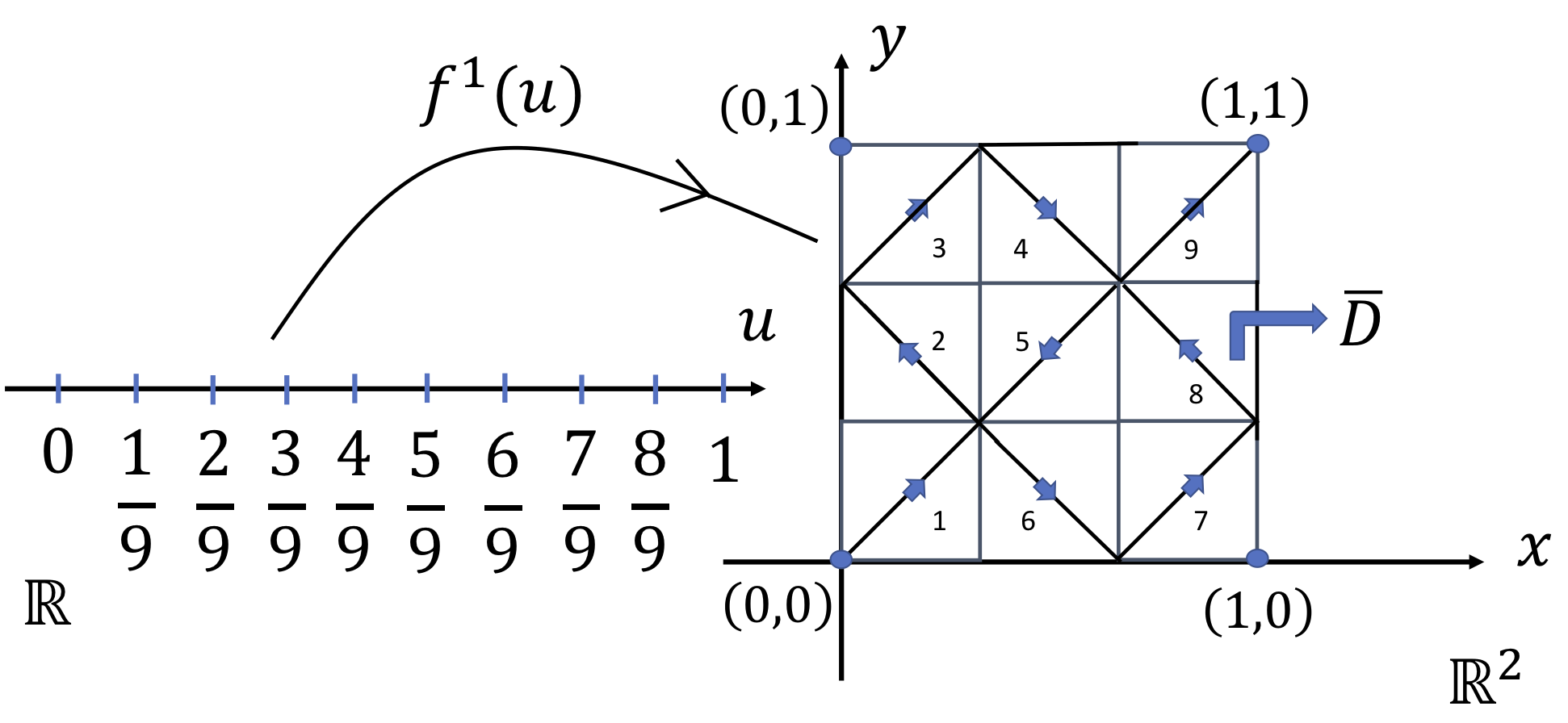}
\end{center}
\caption{The graph of the parametrized curve $f^{1}(u)$.}
\end{figure}
Note that the curve $f^{1}(u)$ is defined over nine closed intervals $\left[ \dfrac{k-1}{9}, \dfrac{k}{9} \right]$ for $k \in \{1,2,\cdots,9\} \subset \mathbb{R}$. The image of the curve of figure 1 is a continuous, piece-wise zigzag, oriented, parametrized curve over a closed square $\bar{D} \subset \mathbb{R}$ of unit area inside the $x-y$ plane $\mathbb{R}^2$. The image of the function $f^{1}(u)$ in figure 1 has for the closed domain$[0,1/9]$ the closed range marked as the first linear segment with an arrow and denoted with a "1". Similarly, the closed domain $[1/9, 2/9]$ is mapped to the linear segment denoted "2". The ten corners of the zig-zag curve are situated at $f^{1}(k/9), k=0,1,...,9$.

\begin{figure}
\begin{center}
\includegraphics[scale=0.35]{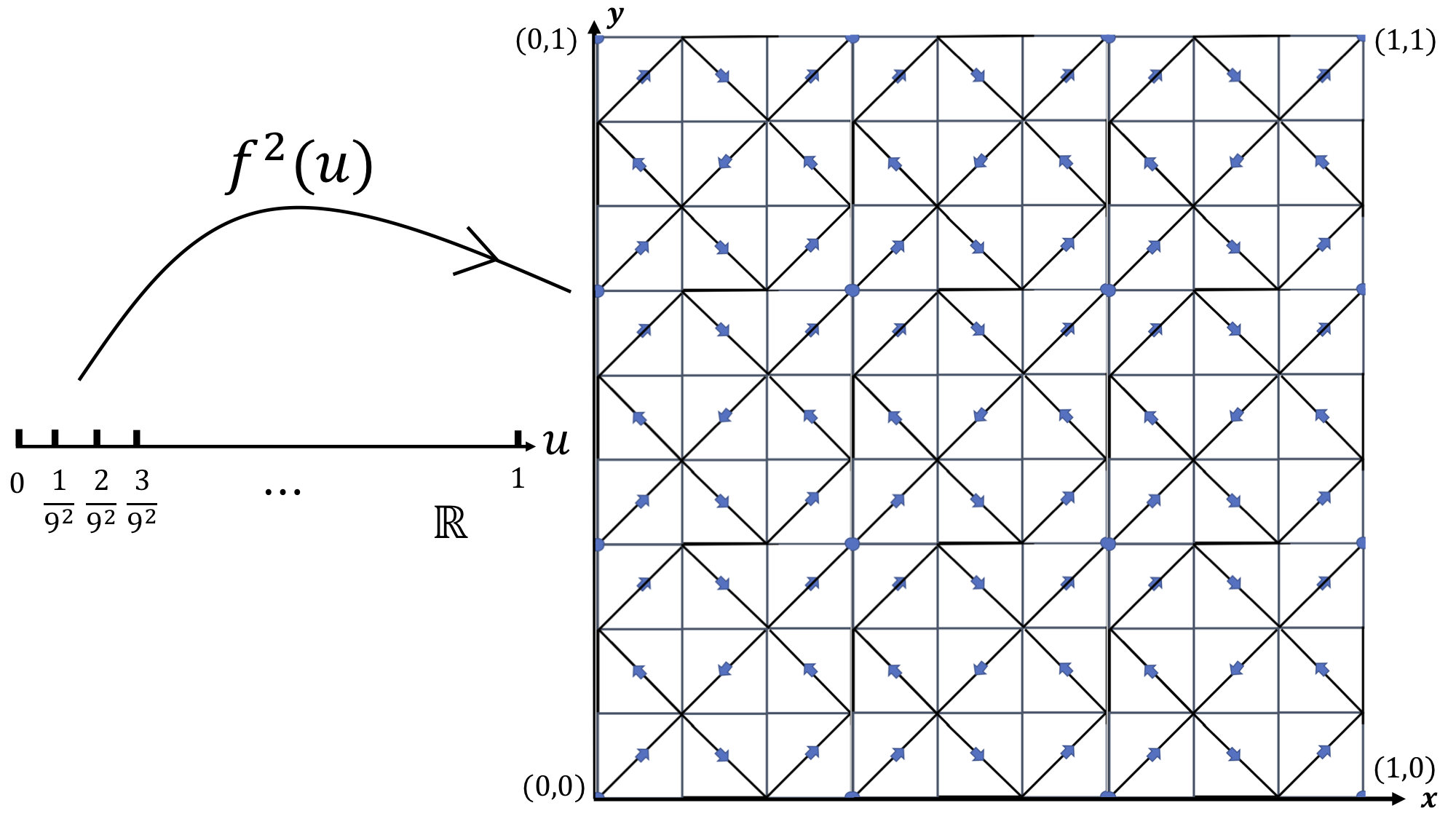}
\end{center}
\caption{The graph of the parametrized curve $f^{2}(u)$.}
\end{figure}

The image of the next parametrized curve $f^{2}(u)$, depicted in figure 2, is continuous, oriented, and piece-wise linear in the $x-y$ plane $\mathbb{R}^2$. It has $9^2 =81$ linear segments inside the unit square $\bar{D}$ within $\mathbb{R}^2$. This figure is constructed by replicating figure 1 nine times within the unit square area resulting in $81$ oriented linear pieces. 

Let us define a distance function $d(f^{1}(u),f^{2}(u))$ abstractly as
\begin{equation}
d(f^{1}(u),f^{2}(u)) := \displaystyle{\textit{sup}_{u \in [0,1]}} ||f^{1}(u)-f^{2}(u)|| \leq \dfrac{\sqrt{2}}{3}.
\end{equation}
Here, the number $\dfrac{\sqrt{2}}{3}$ is the diameter of a square with each side being
$\dfrac{1}{3}$. The sequence of pre-Peano parametrized curves $\{ f^{j}(u)\}^{\infty}_{j=1}$ is obtained by replicating the above process over and over again. Therefore, one may obtain the result that 
\begin{equation}
d(f^{j}(u),f^{j+1}(u)) \leq \dfrac{\sqrt{2}}{3^j}.
\end{equation}
It follows by the uniform Cauchy criterion \cite{Clark} that for each $\epsilon >0$, there exists an integer $N>0$ such that $ d(f^{j}(u),f^{m}(u)) < \epsilon$ for all $j$ and $m \geq N$. Therefore, there exists a uniformly continuous function $f(u)$ such that $\displaystyle{\lim_{j \rightarrow \infty}} f^{j}(u) =f(u)$. Moreover, on can prove that this parametrized curve $f(u)$ \textit{passes through every point of the closed unit square} $\bar{D}$. This is the first example of a Peano curve filling in the area of $\bar{D}$ \cite{Clark}.  

\section{Other Peano curves filling a $(1+1)$-dimensional physical phase space}

We can interpret physically the areas $\bar{D}$ in both figures 1 and 2 as phase cells inside a $(1+1)$-dimensional phase plane. Furthermore, the uncertainty of quantum physics 
\cite{Dirac} $|\Delta x | |\Delta y| \geq 1 $ $(i.e. \, |\Delta q| |\Delta p| \geq 1 , \hbar =1)$ automatically holds inside a phase cell $\bar{D}$ for a quantized particle occupying $\bar{D}$ (here, $\Delta$ stands for an increment and not a finite difference operator). Any possible movement of the quantized particle along the Peano curve $f(u)$ covering $\bar{D}$ is physically unobservable.

Now, we shall define an infinite sequence of functions $\{ h^{M}\}^{\infty}_{M=1}$ from the region $\bar{D} \subset \mathbb{R}^2$ into regions $\{ \bar{D}_{M}\}^{\infty}_{M=1}$ as depicted in figure 3.
\begin{figure}
\begin{center}
\includegraphics[scale=0.35]{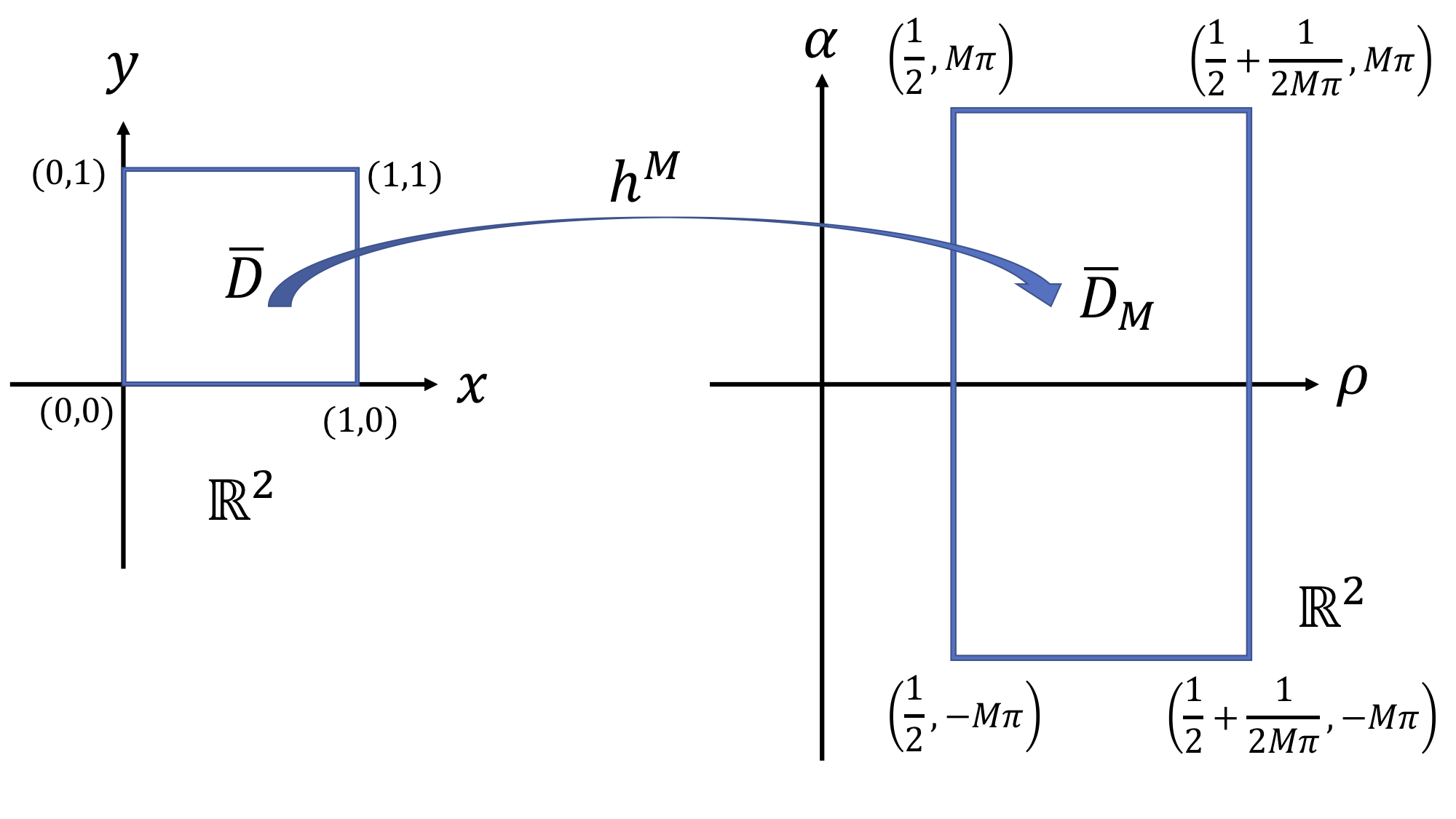}
\end{center}
\caption{The graph of the function $ h^{M}$ for fixed $M$.}
\end{figure}
The function $h^M$ is defined for a specific $M$ as the following linear transformation:
\begin{equation}
\begin{array}{c}
\rho = \left( \dfrac{1}{2M\pi} \right) x +\dfrac{1}{2}, \\
\\
\alpha = (2M \pi) y -M\pi, \\
\\
M \geq 1.
\end{array}
\end{equation}
The Jacobian of these transformations is
\begin{equation}
\dfrac{\partial (\rho, \alpha)}{\partial(x,y)} =1,
\end{equation}
making these transformations canonical transformations of Hamiltonian mechanics \cite{Lanczos,Goldstein}. Therefore, the area of each of the regions $\{ \bar{D}_{M}\}$ is furnished by the double integral of the following  differential 2-form \cite{Spivak,DasTA}
\begin{equation}
\begin{array}{c}
Area (\bar{D}^{M}) :=\displaystyle{{\int \int}_{\bar{D}^{M}}} d\rho \wedge d\alpha = \int_{\frac{1}{2}}^{\frac{1}{2M\pi}  + \frac{1}{2}} \int_{-M\pi}^{M\pi} d\rho d\alpha =1, \\
\\
M \in \{1,2,3,...\}.
\end{array}
\end{equation}

Remarks:

(i) In the limiting process $M \rightarrow \infty$, the sequence of regions $\{ \bar{D}^{M}\}^{\infty}_{M=1}$ collapses into the open vertical line given by $\{(\rho, \alpha) \in \mathbb{R}^2 : \rho =1/2, \alpha \in \mathbb{R} \}$.

(ii) This infinite vertical line is an open string-like phase cell of unit area as seen by using (6) above.

(iii) The infinite vertical line in remark (i) is consistent with the uncertainty principle $|\Delta \rho | |\Delta \alpha | \geq 1 $.

(iv) The string-like phase cell may or may not contain any quanta of a second quantized relativistic wave field (see section 5).

(v) Moreover, unlike in conventional string theory \cite{Green,PolchinskiI,PolchinskiII}, in our approach of string-like phase cells, an open string-like phase cell cannot be of finite length.

Consider another sequence of transformations $\{h^{M}_n\}^{\infty}_{M=1}$ where $n$ is chosen to be a fixed non-negative integer. A typical mapping $h^{M}_n$ maps the closed domain $\bar{D}_M$ into the closed region $\bar{D}_{Mn}$ by the linear transformation
\begin{equation}
\begin{array}{c}
r = \rho + n, \\
\\
\theta = \alpha, \\
\\
\dfrac{\partial (r,\theta)}{\partial (\rho, \alpha)} =1.
\end{array}
\end{equation}
This transformation is exhibited in figure 4. 
\begin{figure}
\begin{center}
\includegraphics[scale=0.35]{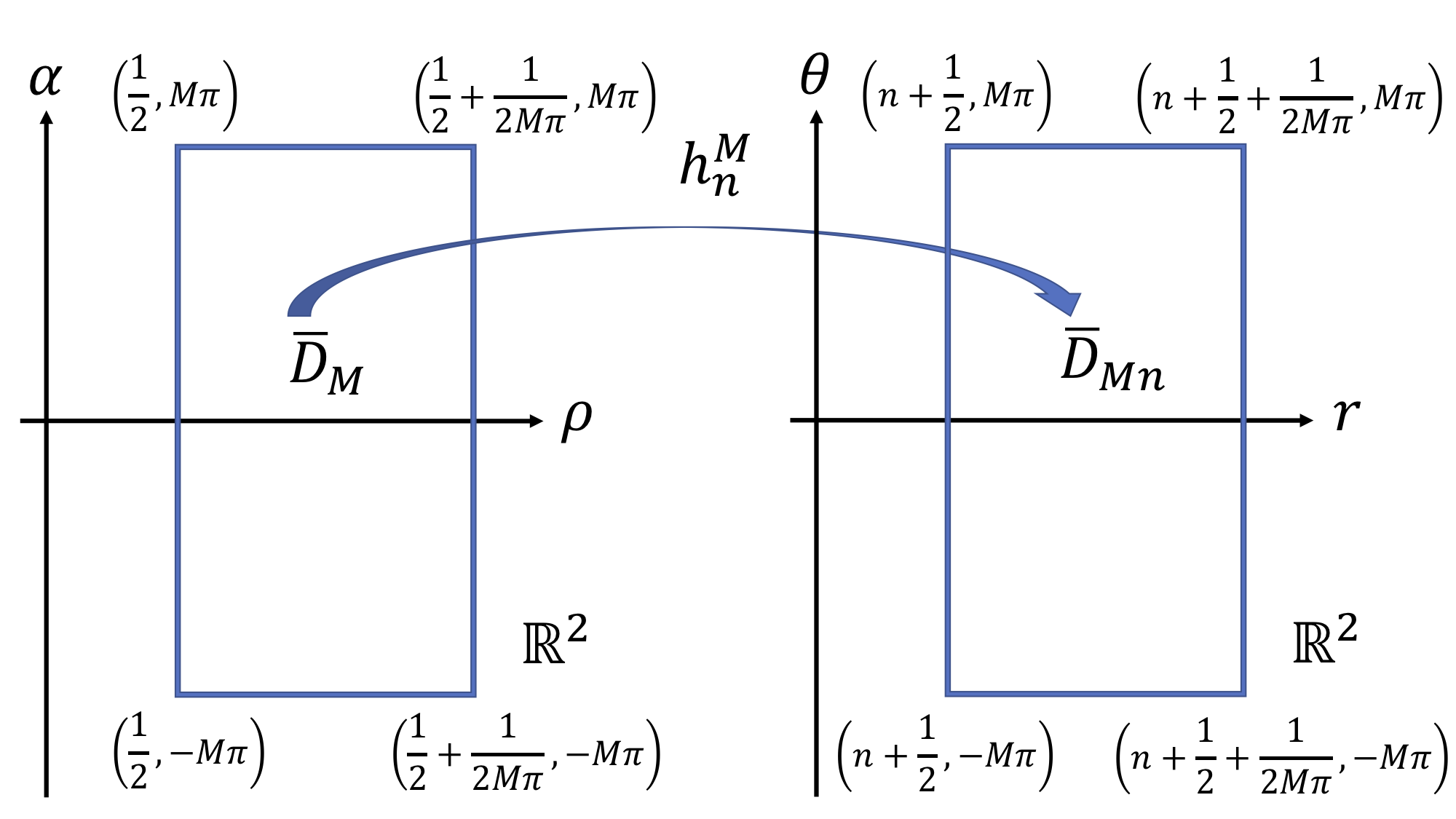}
\end{center}
\caption{The mapping $h^{M}_{n}$ from $\bar{D}_M$ to $\bar{D}_{Mn}$.}
\end{figure}
The area of each domain $\{ \bar{D}_{Mn}\}$ is given by 
\begin{equation}
\begin{array}{c}
Area (\bar{D}_{Mn}) := \displaystyle{{\int \int}_{\bar{D}_{Mn}}} dr \wedge d\theta = \int_{n+\frac{1}{2}}^{n+\frac{1}{2M\pi}  + \frac{1}{2}} \int_{-M\pi}^{M\pi} dr d\theta =1, \\
\\
M \in \{1,2,3,...\}, \,\,\,\ n \in \{0,1,2,....\}.
\end{array}
\end{equation}

Now, consider a third sequence of canonical transformations $\{g^{M}_n\}^{\infty}_{M=1}$ for a fixed $n$ as follows,
\begin{equation}
\begin{array}{c}
q = \sqrt{2\rho} \cos \theta, \\
\\
p = \sqrt{2\rho} \sin \theta, \\
\\
\dfrac{\partial (q,p)}{\partial (\rho, \theta)} =1, \\
\\
q^2 + p^2 = 2r > 0, \\
\\
\left( \dfrac{p}{q} \right) = \tan \theta, \,\,\, q \neq 0.
\end{array}
\end{equation}
The closed domain of each of the mappings of $\{g^{M}_n\}^{\infty}_{M=1}$ is furnished by
\begin{equation}
\bar{D}_{Mn} := \left\{(r,\theta) \in \mathbb{R}^2 :\left( n+\frac{1}{2} \right) \leq r \leq \left( n+\frac{1}{2M\pi}  + \frac{1}{2} \right) , -M\pi \leq \theta \leq M\pi \right\}.
\end{equation}
The corresponding co-domain $\bar{A}_{Mn}$ in the $q-p$ phase plane is provided by the annular region:
\begin{equation}
\begin{array}{c}
\bar{A}_{Mn} := \left\{(q,p) \in \mathbb{R}^2 :2n+1 \leq q^2 +p^2 \leq \left( 2n+1+\frac{1}{M\pi}\right) \right\}, \\
\\
Area (\bar{A}_{Mn}) := \displaystyle{\int\int_{\bar{A}_{Mn}}} dp \wedge dq = 1.
\end{array}
\end{equation}
The mappings $\{g^{M}_n\}^{\infty}_{M=1}$ are depicted in the figure 5.
\begin{figure}
\begin{center}
\includegraphics[scale=0.35]{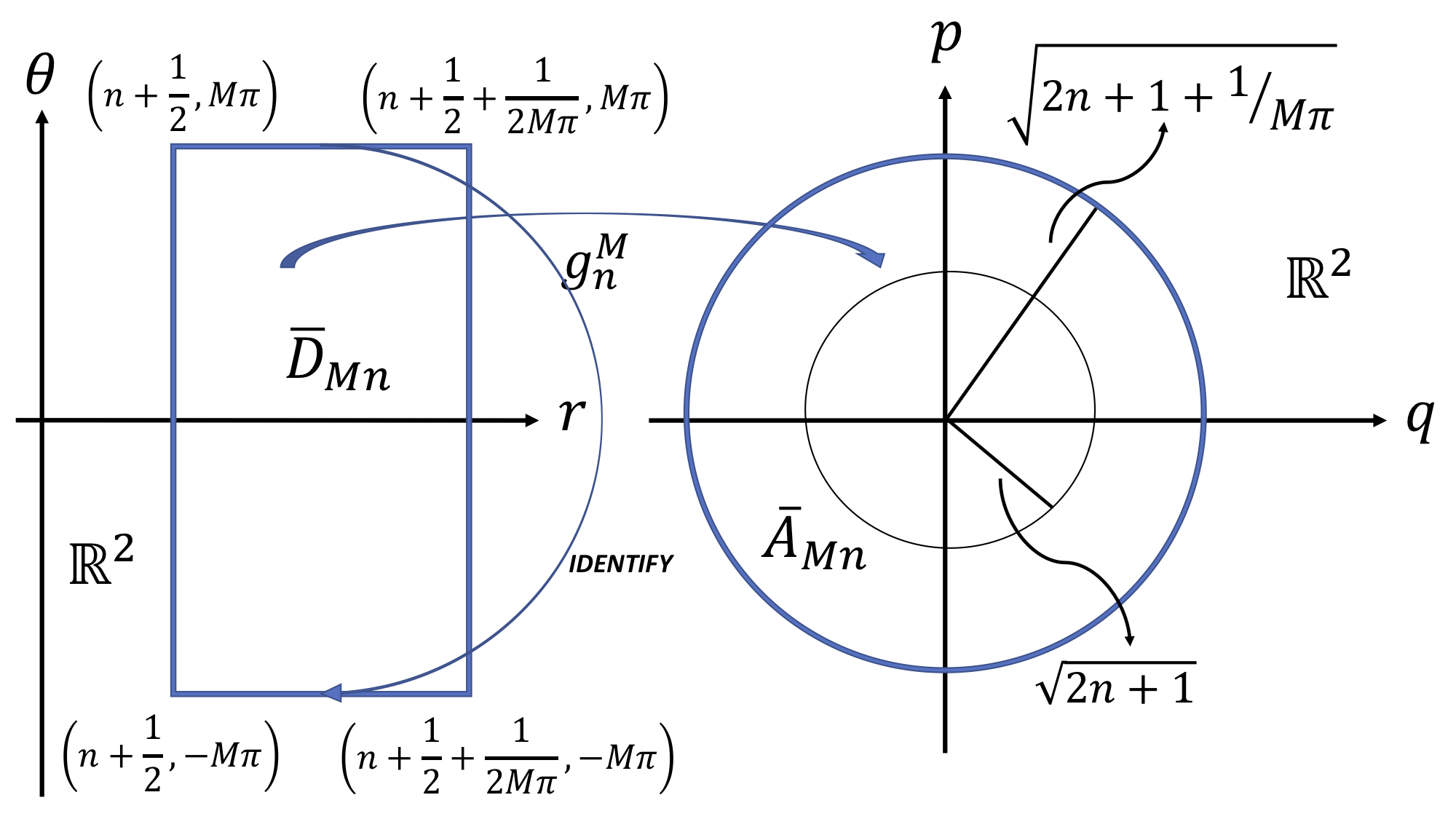}
\end{center}
\caption{The mapping $g^{M}_{n}$ with topological identifications of the two horizontal sides of $\bar{D}_{Mn}$.}
\end{figure}
Note that the annular region $\bar{A}_{Mn}$ is not a simply connected region of the circular disk $q^2 +p^2 = 2n +1 +\frac{1}{M\pi}$. Furthermore, the winding number of each boundary of the  annular region $\bar{A}_{Mn}$ is exactly $M$.

\section{The Peano circle $\bar{S}^{1}_{n}$ in the $(1+1)$-dimensional phase plane}
We shall now investigate the closed domain $\bar{A}_{Mn}$ of the mapping $g^{M}_{n}$ for a fixed $n$ as exhibited in figure 5. Using (9) and (11), the inner and outer boundaries of the closed co-domain $\bar{A}_{Mn}$ are found to be
\begin{equation}
\begin{array}{c}
\partial_{-}[\bar{A}_{Mn}]:= \{(q,p) \in \mathbb{R} : q^2 + p^2 =2n+1 \}, \\
\\
\partial_{+}[\bar{A}_{Mn}]:= \left\{(q,p) \in \mathbb{R} : q^2 + p^2 =2n+1 + \frac{1}{M\pi} \right\}.
\end{array}
\end{equation}
Consider the limiting map $\displaystyle{\lim_{M \rightarrow \infty}} \{g^{M}_{n}\}_{M=1}^{\infty}$. Clearly, one has for this limiting map
\begin{equation}
\displaystyle{\lim_{M \rightarrow \infty}} \left( \left\{(q,p) \in \mathbb{R} : q^2 + p^2 =2n+1 + \frac{1}{M\pi} \right\} \right) = \partial_{-}[\bar{A}_{Mn}]
\end{equation}
which is depicted in figure 6.
\begin{figure}
\begin{center}
\includegraphics[scale=0.35]{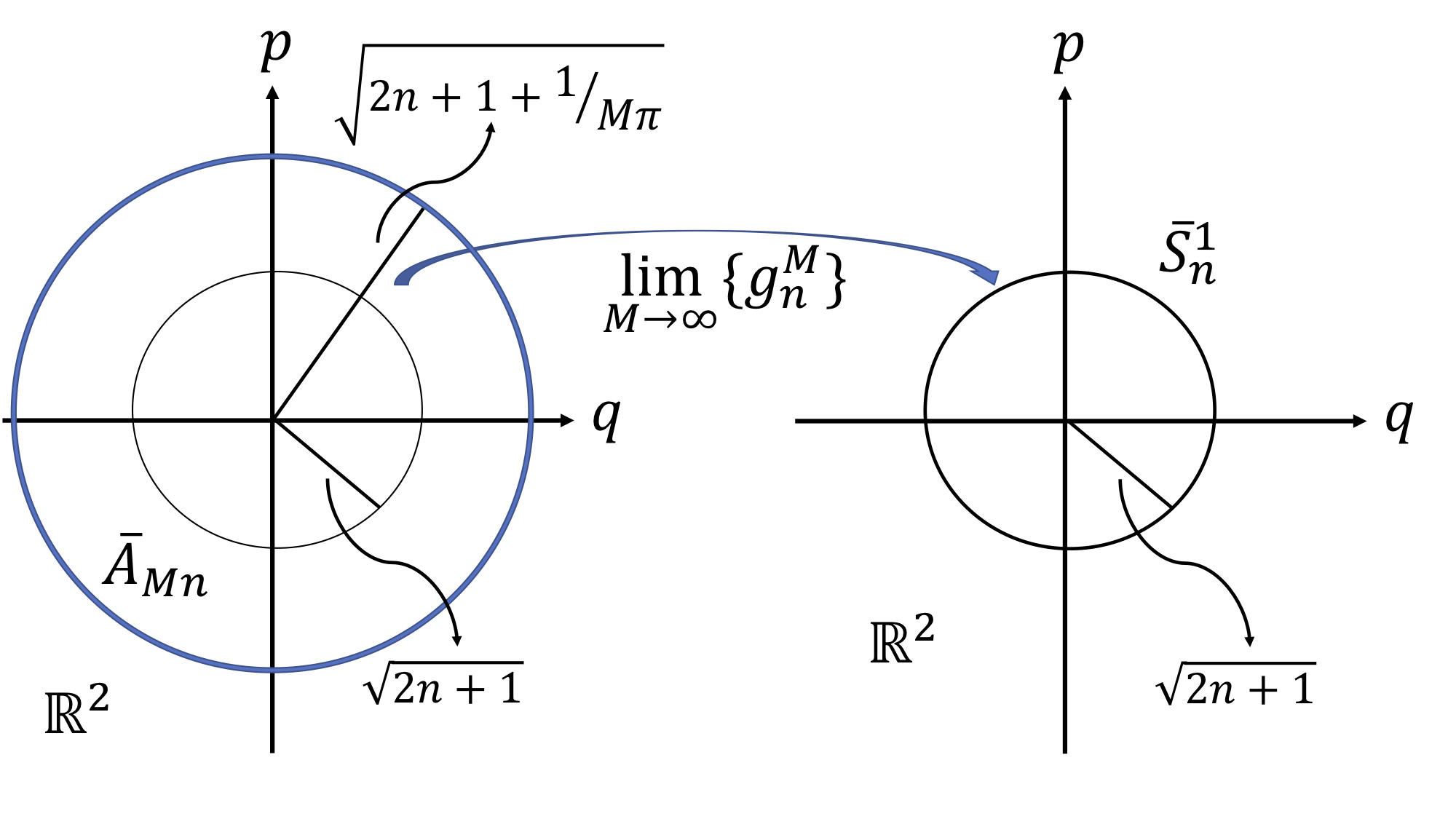}
\end{center}
\caption{The limiting map $\displaystyle{\lim_{M \rightarrow \infty}} \{g^{M}_{n}\}$ and the closed range as the Peano circle $\bar{S}^{1}_{n}$.}
\end{figure}

Remarks:

(i) Note that by (11), $Area (\bar{A}_{Mn})=1$ for all $M \in \{1,2,3,...\}$.

(ii) Since the closed range of the mapping $\displaystyle{\lim_{M \rightarrow \infty}} \{g^{M}_{n}\} =\bar{S}^{1}_{n}$, the Peano circle $\bar{S}^{1}_{n}$ itself is endowed with the unit $(1+1)$-dimensional area.

(iii) The Peano circle can act as a string-like \cite{Green,PolchinskiI,PolchinskiII} phase-cell.

(iv) The quantum mechanical uncertainty principle $|\Delta q | |\Delta p| \geq 1 $ holds for each quanta inside the phase cell $\bar{S}^{1}_{n}$.

(v) The original Peano-curve $\displaystyle{\lim_{j \rightarrow\infty}} f^{j}(u) = f(u)$ is squashed inside the Peano circle $\bar{S}^{1}_{n}$.

(vi) The first quantized oscillator introduced in part \textbf{I} of this work \cite{DasRCI} follows a zig-zag motion or zitter-bewengung \cite{Greiner} along the squashed Peano curve inside $\bar{S}^{1}_{n}$.

(vii) Unlike the Peano circle $\bar{S}^{1}_{n}$ here, the one-dimensional circle of figure 2 in part \textbf{I} \cite{DasRCI} cannot act as a two-dimensional phase cell within the $q-p$ phase plane.

Consider now the infinite $[(1+1)+1]$-dimensional hyper-circular cylinder $\bar{S}^{1}_{n} \times \mathbb{R}$ inside the $[(1+1)+1]$-dimensional discrete phase plane plus continuous time $t \in \mathbb{R}$ called the state space \cite{Lanczos}. Within this state space is a world-sheet like object \cite{Green,PolchinskiI,PolchinskiII} depicted in figure 7.
\begin{figure}
\begin{center}
\includegraphics[scale=0.35]{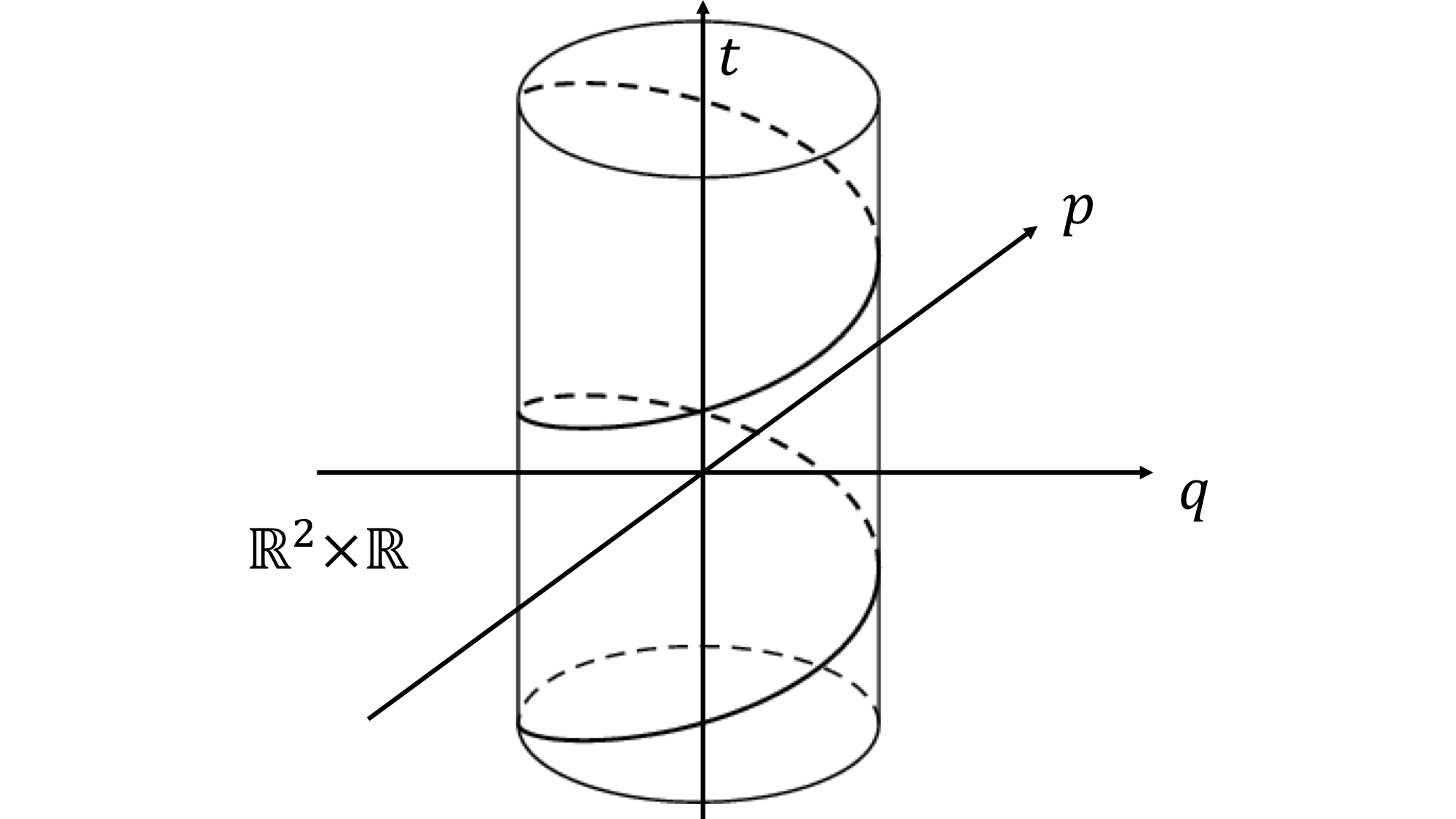}
\end{center}
\caption{The $[(1+1)+1]$-dimensional infinite hyper-cylinder $\bar{S}^{1}_{n} \times \mathbb{R}$ inside the $[(1+1)+1]$-dimensional discrete phase plane plus continuous time.}
\end{figure}
The $[(1+1)+1]$-dimensional  circular cylinder $S^{1}_{n} \times \mathbb{R}$  inside the 
state space in figure 4 in part \textbf{I} \cite{DasRCI} cannot represent an evolving phase cell unlike the hyper-circular cylinder $\bar{S}^{1}_{n} \times \mathbb{R}$  depicted in figure 7.

Let us summarize what we have achieved so far. Figures 1 and 2 have described the area filling curve of Peano, which completely fills the area of a unit square. We identify this unit area as a possible phase cell within a $(p,q)$-phase plane. The area filling Peano curve is identified as the trajectory of one quanta of certain quantized simple harmonic oscillators introduced in \cite{DasRCI}. The trajectory of the Peano curve is not observable due to the uncertainty principle of quantum mechanics \cite{Greiner,Bethe}. Equations (4), (5), (7), and (9) indicate sequences of area preserving mathematical mappings physically representing canonical transformations of Hamiltonian mechanics \cite{Lanczos,Goldstein}. Therefore, in figure 5, the original unit square has yielded a sequence of circular, annular domains each of unit area. In figure 6, this sequence of circular annular domains collapses into a Peano circle $\bar{S}^{1}_{n}$ of unit area. The quanta of the Planck oscillator \cite{DasRCI} still goes around the Peano circle with constant energy $\sqrt{2n+1}$ in a zig-zag pattern but is undetectable by external observations. 

Note that a Peano circle $\bar{S}^{1}_{n}$ of unit area in the $(1+1)$-phase plane is analogous to a closed string of prototypical string theory \cite{Green,PolchinskiI,PolchinskiII}. Figure 7 depicts the $[(1+1)+1]$-dimensional hyper-circular-cylinder inside the three-dimensional state space that is analogous to the two-dimensional world-sheet evolution of closed strings. 

\section{The Klein-Gordon equation in a $[(1+1)+1]$-dimensional discrete phase-plane and continuous time}
Now, we investigate the second quantization of the Klein-Gordon equation in a $[(1+1)+1]$-dimensional discrete phase-plane and continuous time scenario. The relativistic  Klein-Gordon equation in a $[(2+2+2)+1]$-dimensional discrete phase-plane and continuous time was discussed in section 7 of part \textbf{I} of this paper \cite{DasRCI}. The second quantized scalar field linear operator, denoted by $\Phi(n,t)$, acts on a Hilbert space bundle \cite{Choquet,DasTA}. The corresponding Klein-Gordon equation is 
\begin{equation}
\begin{array}{c}
[\mathbf{P} \cdot \mathbf{P} -(\mathbf{P_t})^2 +m^2 \mathbf{I}]\overrightarrow{\mathbf{\Psi}} = \overrightarrow{\mathbf{0}}, \\
or \\
(\Delta^{\#})^2 \Phi(n,t)- (\partial_t)^2 \Phi(n,t) - m^2 \Phi(n,t) = 0.
\end{array}
\end{equation}
Consider the complex valued functions $\xi_n(k)$ involving Hermite polynomials $H_n(k)$ \cite{Olver},
\begin{equation}
\begin{array}{c}
\xi_n(k) := \dfrac{(i)^n e^{-k^2/2} H_n(k)}{\pi^{1/4}2^{n/2}\sqrt{n!}}, \\
\\
\xi_n(-k) = \overline{\xi_n(k)}, \\
\xi_{2n+1}(0)=0, \\
\displaystyle{\int_{\mathbb{R}}} \overline{\xi_m(k)} \xi_n(k) dk = \delta_{mn}, \\
\displaystyle{\sum_{n=0}^{\infty}} \overline{\xi_n(k)} \xi_n(\hat{k}) = \delta(k-\hat{k}), \\
\Delta^{\#} \xi_n(k) = ik \xi_n(k).
\end{array}
\end{equation}
Here, $\delta(k-\hat{k})$ indicates the Dirac delta distribution function \cite{Zemanian}. A special class of exact solutions to the partial difference-differential equations of (14) are given by
\begin{equation}
\begin{array}{c}
\Phi^{-}(n,t) := \displaystyle{\int_{\mathbb{R}}} \dfrac{1}{\sqrt{2 \omega(k)}} \left[ A(k) \xi_n(k) e^{-i \omega t} \right] dk,  \\
\\
\Phi^{+}(n,t) := \displaystyle{\int_{\mathbb{R}}} \dfrac{1}{\sqrt{2 \omega(k)}} \left[ B^{\dagger}(k) \overline{\xi_n(k)} e^{i \omega t} \right] dk, \\
\\
\Phi(n,t) := \Phi^{-}(n,t) + \Phi^{+}(n,t), \\
\\
\omega = \omega(k) := + \sqrt{k^2 +m^2} > 0.
\end{array}
\end{equation}

The Fourier-Hermite integrals in (16) are supposed to be uniformly convergent \cite{Buck}. Moreover, the linear operators $A(k), A^{\dagger}(k), B(k)$ and $B^{\dagger}(k)$ (creation and annihilation operators in momentum space) acting linearly on a Hilbert space bundle must satisfy the following commutation relations:
\begin{equation}
\begin{array}{c}
[A(k),A^{\dagger}(\hat{k})] = [B(k),B^{\dagger}(\hat{k})]=\delta(k-\hat{k})I(k), \\
\\
\, [A(k),A(\hat{k})] = [A^{\dagger}(k), A^{\dagger}(\hat{k})] = [B(k),B(\hat{k})] = [B^{\dagger}(k), B^{\dagger}(\hat{k})] = 0, \\
\\
N^{+}(k) := A^{\dagger}(k) A(k), \\
\\
N^{-}(k) := B^{\dagger}(k) B(k),
\end{array}
\end{equation}
where the eigenvalues of the number operators $N^{\pm}(k)$ are the set $\{0,1,2,...\}$.
Physically speaking, the scalar field operator $\Phi(n,t)$ represents a collection of massive, spin-less, electrically charged, second quantized scalar particle excitations. 

One can show the following relations for total energy, total momentum and total electric charge respectively \cite{DasI,DasII},
\begin{equation}
\begin{array}{c}
\mathcal{H} := \displaystyle{\int_{\mathbb{R}}} [N^{+}(k)+N^{-}(k)+\delta(0)I(k)] \omega(k) dk, \\
\\
\mathcal{P} := \displaystyle{\int_{\mathbb{R}}} [N^{+}(k)+N^{-}(k)] k dk, \\
\\
\mathcal{Q} := e \displaystyle{\int_{\mathbb{R}}} [N^{+}(k)-N^{-}(k)] dk.
\end{array}
\end{equation}
The divergent null point energy term $\delta(0)I(k)$ may be ignored for physical interpretations but cannot be rectified directly.

\section{Fibre Bundles}

Let $M$ and $M^{\#}$ be two topological manifolds \cite{Choquet,DasTA}. The Cartesian product  $M \times M^{\#}$ and the projection mapping $\Pi$ from $M \times M^{\#}$  into $M$ constitute a product or trivial bundle $(M \times M^{\#}, M, \Pi)$ depicted in figure 8. The vertical linear segment inside $M \times M^{\#}$ is called a fibre \cite{Choquet,DasTA}. 
\begin{figure}
\begin{center}
\includegraphics[scale=0.35]{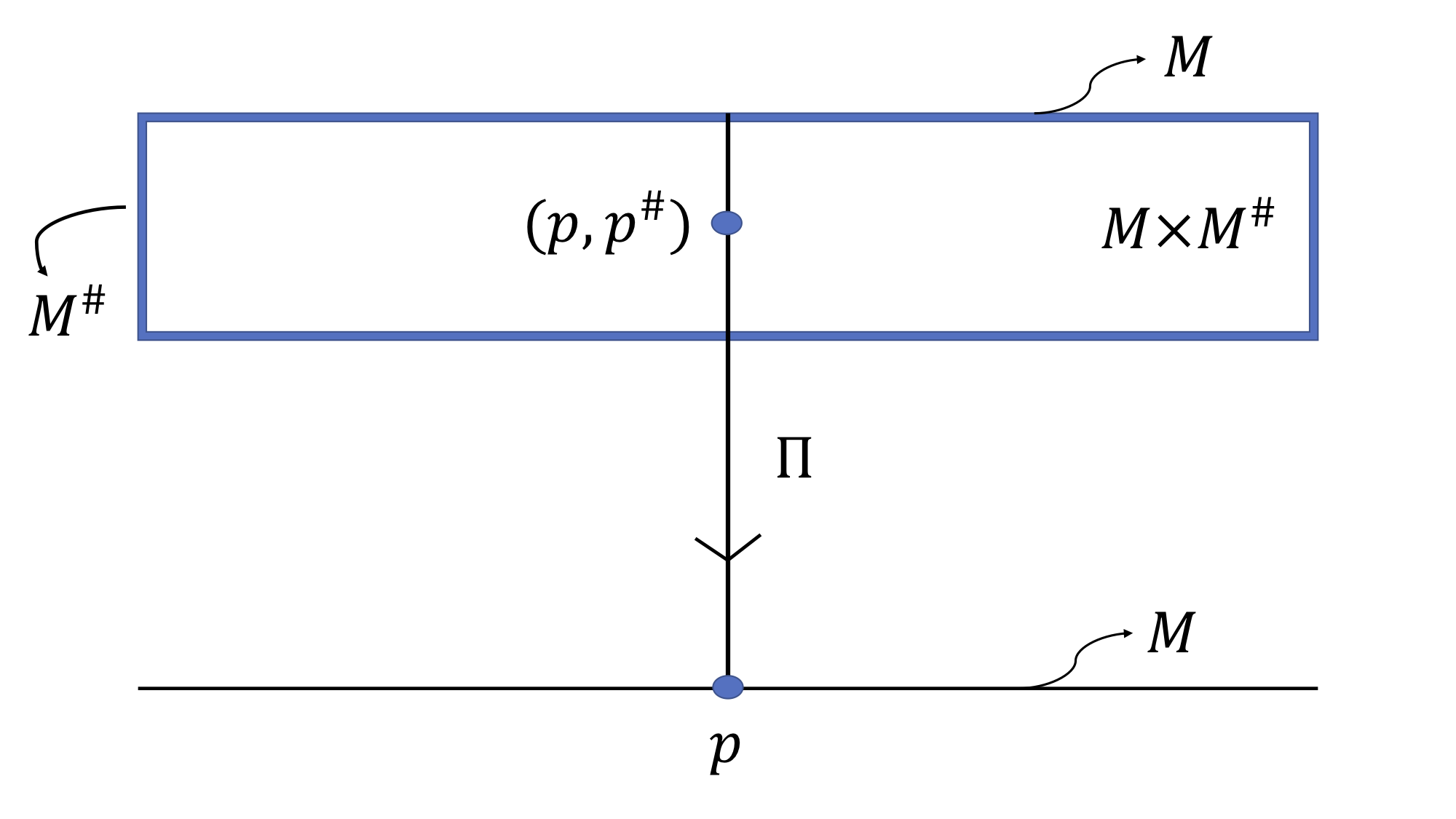}
\end{center}
\caption{The trivial or product bundle $(M \times M^{\#}, M, \Pi)$.}
\end{figure}
Figure 9 provides an explicit example of a product bundle using our Peano cirle $\bar{S}^{1}_n$ to create a vertical circular cylinder $\bar{S}^{1}_n \times I_t$, 
\begin{figure}
\begin{center}
\includegraphics[scale=0.35]{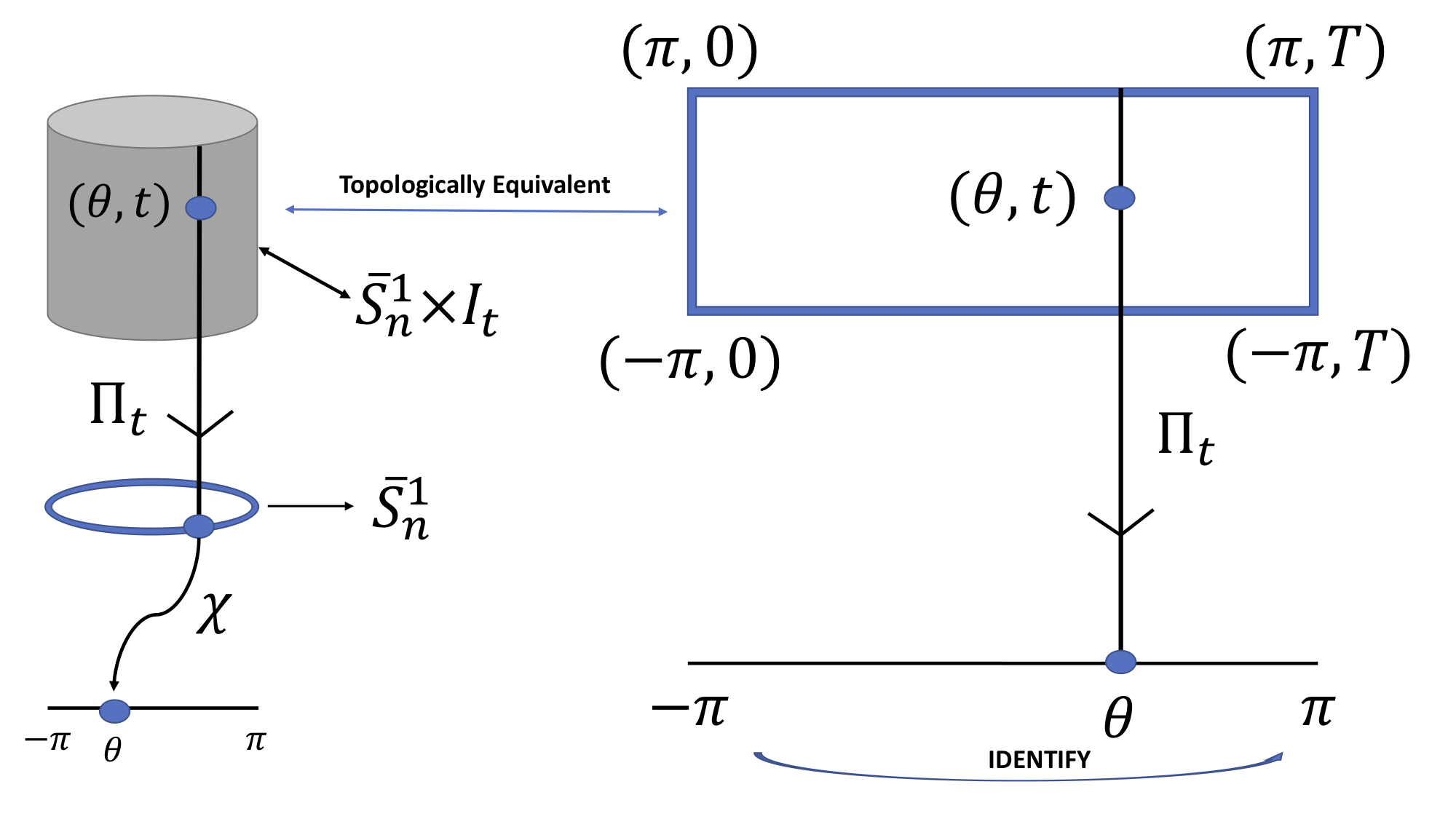}
\end{center}
\caption{The product bundle $(\bar{S}^{1}_n \times I_t, \bar{S}^{1}_n, \Pi_t)$.}
\end{figure}
The closed interval $I_t := [0,T] \subset \mathbb{R}$ and $(\chi, [-\pi, \pi])$ is a coordinate chart \cite{DasTA} in figure 9. The second quantized scalar field operators defined in (16) are restricted for fixed numbers $n$ and $T$ of the product bundle in figure 9.

Another product bundle associated with the linear operators (16) is created by using our Peano circle $\bar{S}^{1}_n$ and the closed linear line segment $I_k := [K_1,K_2] \subset \mathbb{R}$ as in figure 10.
\begin{figure}
\begin{center}
\includegraphics[scale=0.35]{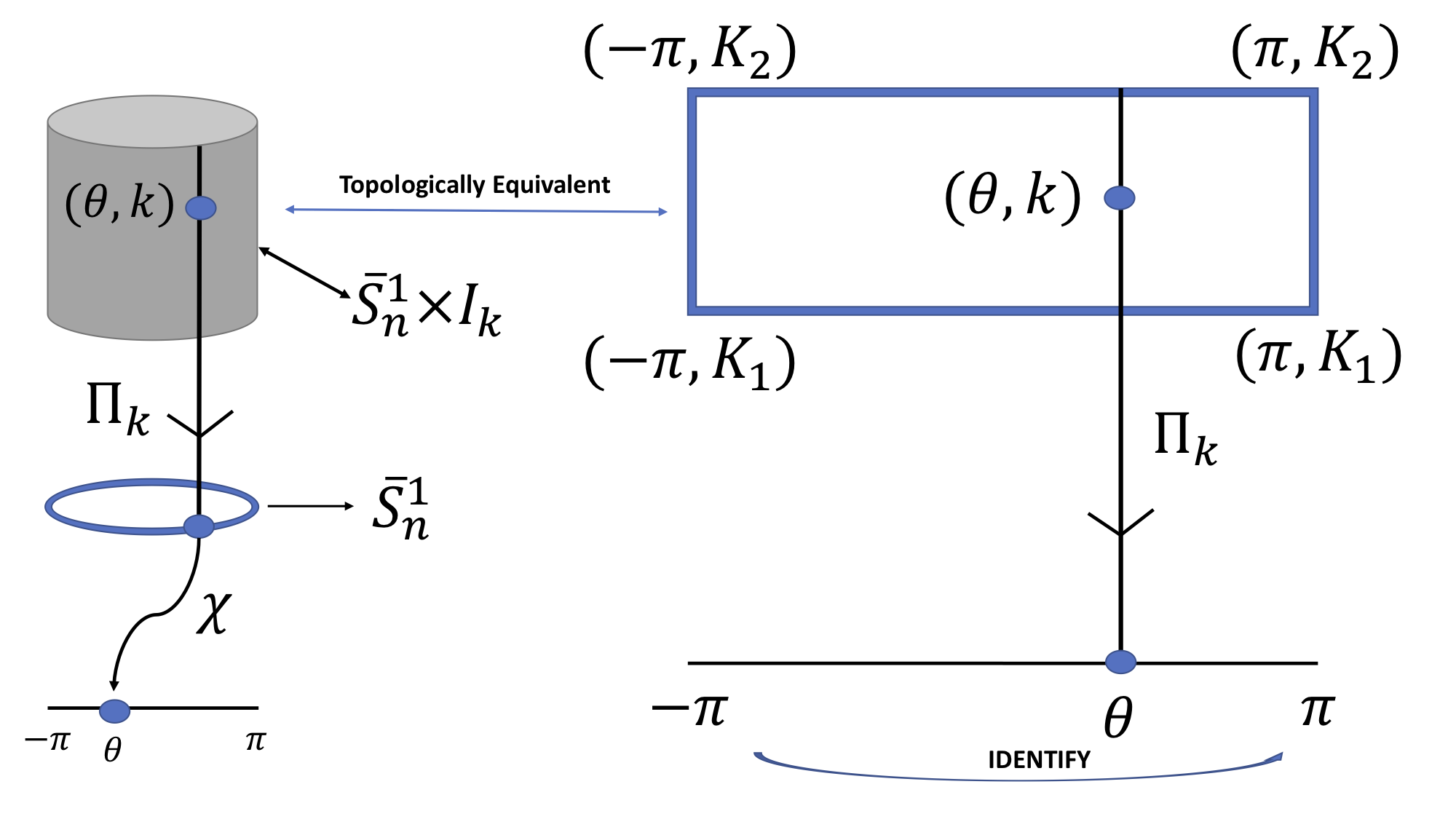}
\end{center}
\caption{The product bundle $(\bar{S}^{1}_n \times I_k, \bar{S}^{1}_n, \Pi_k)$.}
\end{figure}

Remarks:

(i) Figure 9 represents geometrically the occupation of a tiny first quantized Planck oscillator of figures 6 and 7 in the interval $[0,T] \subset \mathbb{R}$ (the full actual physics takes place in the time interval $-\infty< t < \infty$).

(ii) Moreover, in figure 9, the second quantized scalar field $\Phi(n,t)$ quanta can cohabit with a tiny first quantized Planck oscillator during the time interval $I_t := [0,T] \subset \mathbb{R}$.

(iii) Furthermore, in figure 10, the second quantized scalar field quanta with a range of momentum $K_1<k<K_2$ can cohabit with a tiny first quantized Planck oscillator.

(iv) The complete mathematical treatment of this phenomena involves the product bundles $(\bar{S}^{1}_n \times I_t, \bar{S}^{1}_n, \Pi_t) \times (\bar{S}^{1}_n \times I_k, \bar{S}^{1}_n, \Pi_k)$ (which is difficult to depict).

\section{Discrete phase space and hyper-tori like phase cells}
Here, we shall extend the various mappings of figures 3,4, and 5 to the $(2+2+2)$-dimensional discrete phase space arena. Choosing the fixed indices $ a \in \{1,2,3\}$ and $M \in \{1,2,3, \cdots \}$, the relevant mappings are illustrated in figure 11. It depicts the composite map $g^{M}_{n^a} \circ h^{M}_{n^a} \circ h^{M}_{a}$ for fixed indices $a$ and $M$.
\begin{figure}
\begin{center}
\includegraphics[scale=0.30, angle = 270]{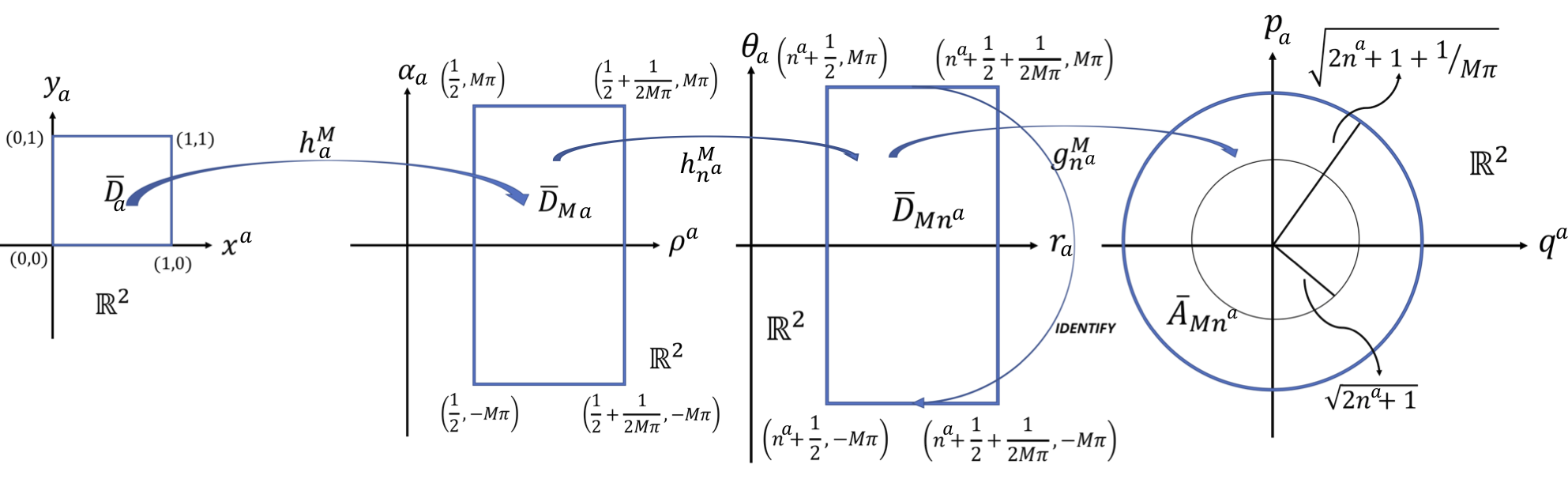}
\end{center}
\caption{The discrete phase space mappings $h^{M}_{a}, h^{M}_{n^a}$, and $g^{M}_{n^a}$.}
\end{figure}

The closed domains and the closed range of the composite map $g^{M}_{n^a} \circ h^{M}_{n^a} \circ h^{M}_{a}$ are provided by
\begin{equation}
\begin{array}{c}
closed \,\,\ domain\,\left[g^{M}_{n^a} \circ h^{M}_{n^a} \circ h^{M}_{a}\right] = \bar{D}_a \subset \mathbb{R}^2, \\
\\
closed \,\,\ range\,\left[g^{M}_{n^a} \circ h^{M}_{n^a} \circ h^{M}_{a}\right] = \bar{A}_{Mn^a} \subset \mathbb{R}^2, \\
or \\
\bar{A}_{Mn^a} = \left[g^{M}_{n^a} \circ h^{M}_{n^a} \circ h^{M}_{a}\right](\bar{D}_a).
\end{array}
\end{equation}
where the last relation illustrates a set theoretic mapping \cite{Goldberg}.

Consider the sequence of composite mappings $\left\{g^{M}_{n^a} \circ h^{M}_{n^a} \circ h^{M}_{a}\right\}_{M=1}^{\infty}$ for a fixed $a$ and $n^a$. The limiting set-theoretic mapping from figure 6 and the last relation of (19) is given by
\begin{equation}
\bar{S}^{1}_n = \displaystyle{\lim_{M\rightarrow \infty}}\left\{\left[g^{M}_{n^a} \circ h^{M}_{n^a} \circ h^{M}_{a}\right](\bar{D}_a)\right\}
\end{equation}
From this, one can derive the set-theoretic Cartesian product mapping \cite{Goldberg,Lightstone}:
\begin{equation}
\begin{array}{c}
\bar{S}^{1}_{n^1} \times \bar{S}^{1}_{n^2} \times \bar{S}^{1}_{n^3}= \displaystyle{\lim_{M\rightarrow \infty}} \left( \left\{ \bar{A}_{Mn^1} \right\} \times \left\{ \bar{A}_{Mn^2} \right\} \times \left\{ \bar{A}_{Mn^3} \right\} \right) , \\
or   \\
\bar{S}^{1}_{n^1} \times \bar{S}^{1}_{n^2} \times \bar{S}^{1}_{n^3}= 
\displaystyle{\lim_{M\rightarrow \infty}} \left( \left\{\left[g^{M}_{n^1} \circ h^{M}_{n^1} \circ h^{M}_{1}\right](\bar{D}_1)\right\} \times \left\{\left[g^{M}_{n^2} \circ h^{M}_{n^2} \circ h^{M}_{2}\right](\bar{D}_2)\right\} \right. \\
\left. \times \left\{\left[g^{M}_{n^3} \circ h^{M}_{n^3} \circ h^{M}_{3}\right](\bar{D}_3)\right\}  \right).
\end{array}
\end{equation}
Furthermore, one has
\begin{equation}
\begin{array}{c}
\left[g^{M}_{n^1} \times h^{M}_{n^1} \times h^{M}_{1} \right] \times \left[g^{M}_{n^2} \times h^{M}_{n^2} \times h^{M}_{2}\right] \times \left[g^{M}_{n^3} \times h^{M}_{n^3} \times h^{M}_{3} \right] \left(\bar{D}_1 \times \bar{D}_2 \times \bar{D}_3 \right)  :=  \\
\\
 \left\{\left[g^{M}_{n^1} \circ h^{M}_{n^1} \circ h^{M}_{1}\right](\bar{D}_1)\right\} \times \left\{\left[g^{M}_{n^2} \circ h^{M}_{n^2} \circ h^{M}_{2}\right](\bar{D}_2)\right\}
\times \left\{\left[g^{M}_{n^3} \circ h^{M}_{n^3} \circ h^{M}_{3}\right](\bar{D}_3)\right\}  
\end{array}
\end{equation}
From the above two relations, a new set-theoretic mapping is furnished by 
\begin{equation}
\begin{array}{c}
\bar{S}^{1}_{n^1} \times \bar{S}^{1}_{n^2} \times \bar{S}^{1}_{n^3}= \displaystyle{\lim_{M\rightarrow \infty}} 
\left\{\left[g^{M}_{n^1} \times h^{M}_{n^1} \times h^{M}_{1} \right] \times \left[g^{M}_{n^2} \times h^{M}_{n^2} \times h^{M}_{2}\right] \times \right. \\ 
\\
\left.\left[ g^{M}_{n^3}  \times h^{M}_{n^3} \times h^{M}_{3} \right] \left(\bar{D}_1 \times \bar{D}_2 \times \bar{D}_3 \right) \right\} 
\subset \mathbb{R}^2 \times \mathbb{R}^2 \times \mathbb{R}^2.
\end{array}
\end{equation}

Remarks:

(i) The $(2+2+2)$-dimensional hyper-sphere or hyper-torus  \cite{Massey} $\bar{S}^{1}_{n^1} \times \bar{S}^{1}_{n^2} \times \bar{S}^{1}_{n^3}$ for three fixed integers $(n^1,n^2,n^3)$ denotes a region where a tiny first quantized Planck oscillator inhabits this region for all time.

(ii) The whole of the  $(2+2+2)$-dimensional discrete phase space contains a denumerably infinite number of concentric hyper-tori each containing a single tiny first quantized Planck oscillator.

(iii) Each $\bar{S}^{1}_{n^1} \times \bar{S}^{1}_{n^2} \times \bar{S}^{1}_{n^3}$ constitutes a phase cell of physical dimension $\left[ \dfrac{ML^2}{T} \right]^3$. Therefore, in the fundamental physical units, each of these phase cells is endowed with a hyper-volume of $ 1 \left[ \dfrac{ML^2}{T} \right]^3$.

(iv) Each of these phase-cells bears a resemblance to a $D$-dimensional hyper-torus in standard string theory \cite{Green,PolchinskiI,PolchinskiII}.

(v) In the arena of $[(2+2+2)+1]$-dimensional discrete phase space and continuous time, the hyper-torus $\bar{S}^{1}_{n^1} \times \bar{S}^{1}_{n^2} \times \bar{S}^{1}_{n^3}$ sweeps out a world-sheet like vertical, circular hyper-cylinder. Such a vertical, circular cylinder always contains just one first quantized tiny Planck oscillator for all time.

(vi) Finally, one or more second quantized scalar field $\Phi(\mathbf{n};t)$ quanta may cohabit the tiny first quantized Planck oscillator temporarily or forever.

\section{Second quantized, relativistic Klein-Gordon field equations in $[(2+2+2)+1]$-dimensional discrete phase space and continuous time}

From part \textbf{I} of this work \cite{DasRCI}, the second quantized, relativistic Klein-Gordon field equations are given by
\begin{equation}
\begin{array}{c}
a,b \in \{1,2,3 \}, \\
\\
n^a \in \{0,1,2,...\},\\
\\
\mathbf{n}:=(n^1, n^2, n^3),\\
\\
\left[(\delta^{ab} \Delta^{\#}_{a} \Delta^{\#}_{b}) -(\partial_t)^2 -m^2 \right] \Phi(\mathbf{n} ;t) =\mathbf{0}.
\end{array}
\end{equation}
The three dimensional extensions of the functions $\xi_{n}(k)$ in (15) imply that \cite{Olver}
\begin{equation}
\begin{array}{c}
\xi_{n^a}(k_a) := \dfrac{(i)^{n^{a}} e^{-k_{a}^2/2} H_{n^{a}}(k_{a})}{\pi^{1/4}2^{n^{a}/2}\sqrt{n^{a}!}}, \\
\\
\mathbf{k} := (k_1, k_2, k_3), \\
\\
\Delta^{\#}_{a} \xi_{n^a}(k_a) = i k_a \xi_{n^a}(k_a), \\
\\
\displaystyle{\sum_{n^1=0}^{\infty} \sum_{n^2=0}^{\infty} \sum_{n^3=0}^{\infty}} \left[\overline{\xi_{n^1}}(k_1) \xi_{n^1}(\hat{k}_1)\right]\left[\overline{\xi_{n^2}}(k_2) \xi_{n^2}(\hat{k}_2)\right]\left[\overline{\xi_{n^3}}(k_3) \xi_{n^3}(\hat{k}_3)\right] \\
=\delta({k_1}-{\hat{k}_1})\delta({k_2}-{\hat{k}_2})\delta({k_3}-{\hat{k}_3})=: \delta^3(\mathbf{k}-\mathbf{\hat{k}}).
\end{array}
\end{equation}

A special class of exact solutions for these relativistic partial difference-differential operator equations (24) are given by
\begin{equation}
\begin{array}{c}
\Phi^{-}(\mathbf{n};t) := \displaystyle{\int_{\mathbb{R}^3}} \dfrac{1}{\sqrt{2 \omega(\mathbf{k})}} \left[ A(\mathbf{k}) \xi_{n^1}(k_1) \xi_{n^2}(k_2) \xi_{n^3}(k_3) e^{-i \omega t} \right] dk_1 dk_2 dk_3,  \\
\\
\Phi^{+}(\mathbf{n};t) := \displaystyle{\int_{\mathbb{R}^3}} \dfrac{1}{\sqrt{2 \omega(\mathbf{k})}} \left[ B^{\dagger}(\mathbf{k}) \overline{\xi_{n^1}}(k_1) \overline{\xi_{n^2}}(k_2) \overline{\xi_{n^3}}(k_3) e^{i \omega t} \right] dk_1 dk_2 dk_3, \\
\\
\Phi(\mathbf{n};t) := \Phi^{-}(\mathbf{n};t) + \Phi^{+}(\mathbf{n};t), \\
\\
\omega = \omega(\mathbf{k}) := + \sqrt{k_1^2 +k_3^2 +k_3^2 +m^2} > 0.
\end{array}
\end{equation}
The Fourier-Hermite integrals in (26) are supposed to be uniformly convergent \cite{Buck}. Moreover, the linear operators $A(\mathbf{k}), A^{\dagger}(\mathbf{k}), B(\mathbf{k})$ and $B^{\dagger}(\mathbf{k})$ (creation and annihilation operators in momentum space) acting linearly on a Hilbert space bundle \cite{Choquet,DasTA} must satisfy the following commutation relations:
\begin{equation}
\begin{array}{c}
[A(\mathbf{k}),A^{\dagger}(\hat{\mathbf{k}})] = [B(\mathbf{k}),B^{\dagger}(\hat{\mathbf{k}})]=\delta(\mathbf{k}-\hat{\mathbf{k}}) \mathbf{I}(\mathbf{k}), \\
\\
\, [A(\mathbf{k}),A(\hat{\mathbf{k}})] = [A^{\dagger}(\mathbf{k}), A^{\dagger}(\hat{\mathbf{k}})] = [B(\mathbf{k}),B(\hat{\mathbf{k}})] = [B^{\dagger}(\mathbf{k}), B^{\dagger}(\hat{\mathbf{k}})] = \mathbf{0}, \\
\\
N^{+}(\hat{\mathbf{k}}) := A^{\dagger}(\mathbf{k}) A(\mathbf{k}), \\
\\
N^{-}(\hat{\mathbf{k}}) := B^{\dagger}(\mathbf{k}) B(\mathbf{k}),
\end{array}
\end{equation}
where the eigenvalues of the number operators $N^{\pm}(\hat{\mathbf{k}})$ are the set $\{0,1,2,...\}$.
The second quantized scalar field operator $\Phi(\mathbf{n};t)$ represents a collection of massive, spin-less, electrically charged, physical particles or quantas. 

One can show the following relations for total energy, total momentum and total electric charge respectively \cite{DasI,DasII},
\begin{equation}
\begin{array}{c}
\mathcal{H} := \displaystyle{\int_{\mathbb{R}^3}} [N^{+}(\mathbf{k})+N^{-}(\mathbf{k})+\delta^3(\mathbf{0})\mathbf{I}(\mathbf{k})] \omega(\mathbf{k}) dk_1 dk_2 dk_3, \\
\\
\mathcal{P}_j := \displaystyle{\int_{\mathbb{R}^3}} [N^{+}(\mathbf{k})+N^{-}(\mathbf{k})] k_j dk_1 dk_2 dk_3, \\
\\
\mathcal{Q} := e \displaystyle{\int_{\mathbb{R}^3}} [N^{+}(\mathbf{k})-N^{-}(\mathbf{k})] dk_1 dk_2 dk_3.
\end{array}
\end{equation}
The divergent null point energy term $\delta^3(\mathbf{0})\mathbf{I}(\mathbf{k})$ may be ignored for physical interpretations but cannot be directly remedied.

\section{Many-particle systems, $(2+2+2)N$-dimensional phase space and $(2+2+2)N$-dimensional hyper-tori as phase cells}

Recall Hamilton's canonical equations of motion for  non-relativistic classical mechanics \cite{Lanczos,Goldstein}
\begin{equation}
\begin{array}{c}
A \in \{1,2,3,...,N\}, \\
\\
\dot{q}^A := \dfrac{d\mathcal{Q}^A(t)}{dt}, \\
\\
\dot{p}_A := \dfrac{d\mathcal{P}_A(t)}{dt}, \\
\\
\dot{q}^A := \dfrac{\partial H}{\partial p_A}(q^1, \cdots , q^{3N}; p_1, \cdots , p_{3N}; t), \\
\\
\dot{p}_A := -\dfrac{\partial H}{\partial q^A}(q^1, \cdots , q^{3N}; p_1, \cdots , p_{3N}; t).
\end{array}
\end{equation}
The corresponding Schr\"{o}dinger wave equation for the first quantized physical system is furnished by
\begin{equation}
\begin{array}{c}
i \dfrac{\partial}{\partial_t} \psi\left(q^1, \cdots , q^{3N};t\right) = \\
H\left(q^1, \cdots , q^{3N}; -i \frac{\partial}{\partial q^1}, \cdots, -i \frac{\partial}{\partial q^{3N}};t\right) \psi(q^1, \cdots , q^{3N};t),
\end{array}
\end{equation}
which has become ubiquitous with standard quantum mechanical systems being verified experimentally for a multitude of specific systems.

From part \textbf{I} of this paper \cite{DasRCI}, the $N=1$ relativistic Klein-Gordon equation in four-dimensional space is given by ($\eta_{\mu \nu}:=[1,1,1,-1]$)
\begin{equation}
\begin{array}{c}
\eta^{\mu \nu} \dfrac{\partial ^2}{\partial q^{\mu}\partial q^{\mu}} \psi\left(q^1, q^2, q^3, q^{4}\right) - m^2 \psi\left(q^1, q^2, q^3, q^{4}\right) = 0,\\
or \\
\delta^{a b} \dfrac{\partial ^2}{\partial q^{a}\partial q^{b}} \psi\left(\mathbf{q};t\right) -  (\partial_t)^2 \psi\left(\mathbf{q};t\right)- m^2 \psi\left(\mathbf{q};t\right) = 0.\\
\end{array}
\end{equation}
The relativistic wave equation for the case of two spin-$\frac{1}{2}$ particles $q_{(1)}^{\mu}$ and $q_{(2)}^{\nu}$ is furnished by the Bethe-Salpeter equation \cite{Bethe}
\begin{equation}
\begin{array}{c}
\left\{\gamma_{(1)}^{\mu} \left[ \left( \dfrac{m_1}{m_1 +m_2} P_{\mu} + P_{\mu}\right) - im_1 \right] \right\} \left\{\gamma_{(2)}^{\nu} \left[ \left( \dfrac{m_1}{m_1 +m_2} P_{\nu} + P_{\nu}\right) - im_2 \right] \right\} \\
\\
\cdot  \psi\left(q_{(1)}^1 - q_{(2)}^1, q_{(1)}^2 - q_{(2)}^2, q_{(1)}^3 - q_{(2)}^3,q_{(1)}^4 - q_{(2)}^4 \right)  = \\
\\
i \bar{G}\left(q_{(1)}^1 - q_{(2)}^1, q_{(1)}^2 - q_{(2)}^2, q_{(1)}^3 - q_{(2)}^3,q_{(1)}^4 - q_{(2)}^4 \right) \psi\left(q_{(1)}^1 - q_{(2)}^1, \cdots \right)
\end{array}
\end{equation}
where $\bar{G}$ is the appropriate Green's function.

In $N+1$-dimensional state space, the generalization  to the relativistic Klein-Gordon wave equation is 
\begin{equation}
\begin{array}{c}
\left[ \delta^{AB} \dfrac{\partial ^2}{\partial q^{A}\partial q^{B}} -(\partial_t)^2 -   m^2 \right] \psi\left(q^1, \cdots , q^N; t \right) = 0,\\
A,B \in \{1, \cdots, N\}.\\
\end{array}
\end{equation}
with a group invariance of $\mathcal{I}[O(N,1)]_+^+$.

Similarly, we can express within a $[(2+2+2)N+1]$-dimensional discrete phase space and continuous time arena, the first quantized partial differential-difference Klein-Gordon equation as 
\begin{equation}
\left[\delta^{AB} \Delta^{\#}_{A} \Delta^{\#}_{B} -(\partial_t)^2 -m^2 \right] \phi(n^1, \cdots , n^N ;t) =0.
\end{equation}
The second quantized version of this generalized Klein-Gordon equation is given by
\begin{equation}
\left[\delta^{AB} \Delta^{\#}_{A} \Delta^{\#}_{B} -(\partial_t)^2 -m^2 \right] \Phi(n^1, \cdots , n^N ;t) =\mathbf{0}
\end{equation}
The group invariance of the above equation is provided by $\mathcal{I}[O(N,1)]_+^+$ (see \cite{DasRCI}).

A special class of exact solutions for these relativistic partial difference-differential operator equations (35) are given by
\begin{equation}
\begin{array}{c}
\Phi^{-}(n^1, \cdots , n^N ;t) := \displaystyle{\int_{\mathbb{R}^N}} \dfrac{1}{\sqrt{2 \omega(\mathbf{k})}} \left[ A(\mathbf{k}) \xi_{n^1}(k_1) \cdots \xi_{n^N}(k_N) e^{-i \omega t} \right] dk_1 \cdots dk_N,  \\
\\
\Phi^{+}(n^1, \cdots , n^N ;t) := \displaystyle{\int_{\mathbb{R}^N}} \dfrac{1}{\sqrt{2 \omega(\mathbf{k})}} \left[ B^{\dagger}(\mathbf{k}) \overline{\xi_{n^1}}(k_1) \cdots \overline{\xi_{n^N}}(k_N) e^{i \omega t} \right] dk_1 \cdots dk_N, \\
\\
\Phi(n^1, \cdots , n^N ;t) := \Phi^{-}(n^1, \cdots , n^N ;t) + \Phi^{+}(n^1, \cdots , n^N ;t), \\
\\
\omega = \omega(k_1, \cdots , k_N) := + \sqrt{\delta^{AB}k_A k_B +m^2} > 0.
\end{array}
\end{equation}
The second quantized linear operators $A(k_1, \cdots , k_N), B^{\dagger}(k_1, \cdots , k_N)$, etc. (creation and annihilation operators in momentum space) acting linearly on a Hilbert space bundle must satisfy very similar commutation relations as those of (27). The linear operator $ \Phi(n^1, \cdots , n^N ;t)$ is defined over the geometric configuration $\bar{S}_{n^1}^1 \times \bar{S}_{n^2}^1 \times \cdots \times \bar{S}_{n^N}^1 \times \mathbb{R}$ where  $\bar{S}_{n^A}^1 $ is a Peano circle of physical dimension $\left[\dfrac{ML^2}{T} \right]$.
\begin{table}
\begin{center}
\includegraphics[scale=0.3]{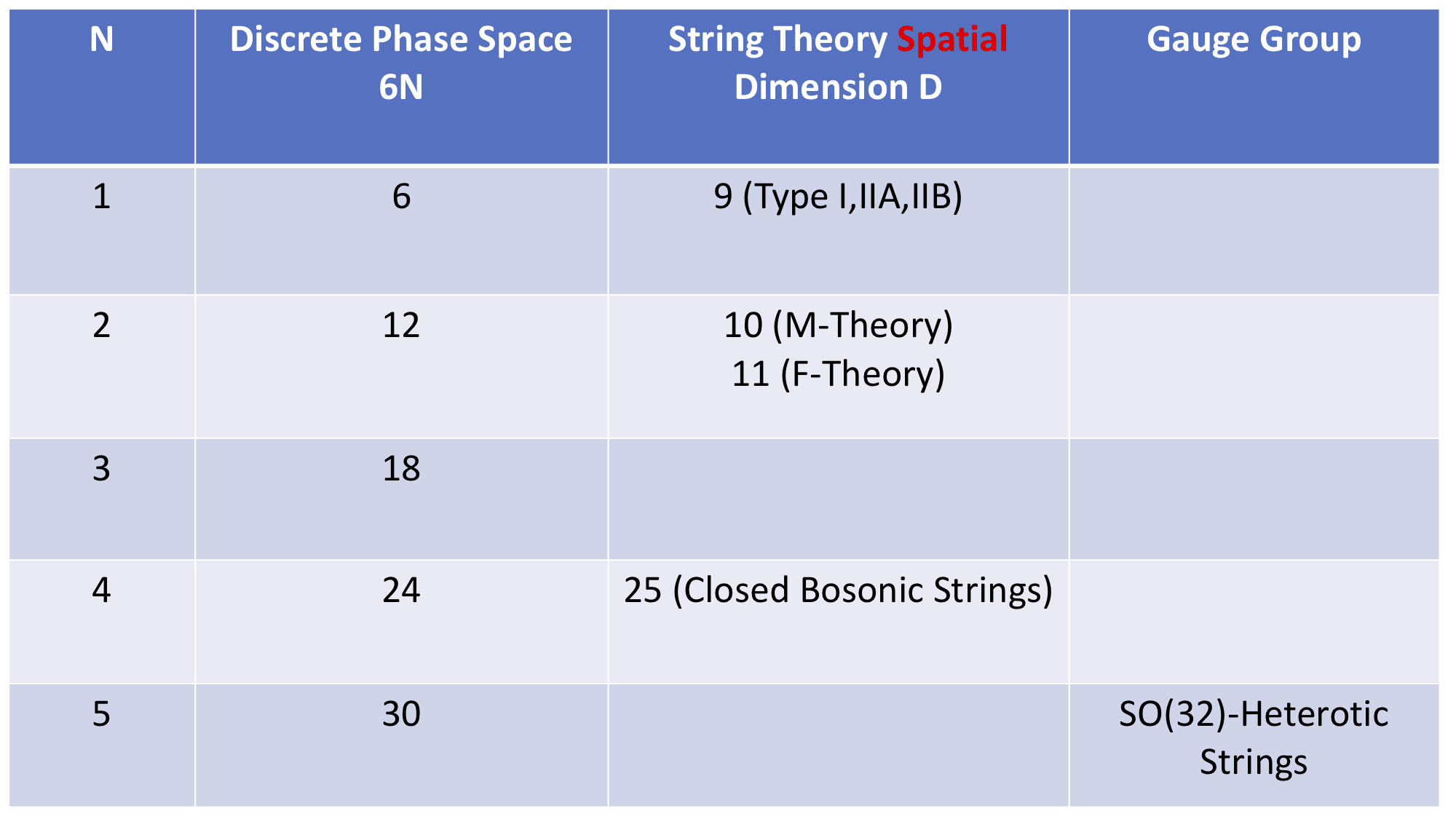}
\end{center}
\caption{Discrete Phase Space and Popular String Theory Dimension Comparison}
\end{table}
 
Finally, we would like to make a final comparison with our method to two popular approaches to quantizing gravity: string theory and loop quantum gravity. Table 1 consists of a higher  dimensional comparison between our discrete phase space theory put forward here and that of various popular string theories \cite{Green,PolchinskiI,PolchinskiII}. There appears to be a striking similarity between the spatial or gauge group dimension of string theory and the dimension of our discrete phase space. One clear advantage of our method is that we have gone further in understanding the scattering matrix of our theory, the $S^{\#}$-matrix, as delineated in \cite{DasI,DasII,DasIII}. In \cite{DasRC}, we have explicitly calculated new Feynman rules for computations of the $S^{\#}$-matrix elements and have derived an exact singularity free Coulomb-type potential within our discrete phase space approach.   See \cite{Green,PolchinskiI,PolchinskiII} for further information on the actual calculation abilities of string theory.

Loop quantum gravity \cite{Rovelli}, and its spin network structure of space-time composed of extremely fine but finite loops has a similarity to our Peano circle $\bar{S}^{1}_{n}$ of unit area in the $(1+1)$ dimensional phase plane. The area filling Peano curve is identified as the trajectory of one quanta as shown here and in \cite{DasRCI}. It would be interesting to further investigate this analogy to possibly find a quantum theory of gravity within our framework or find Peano type motion within the framework of loop quantum gravity.

\section{Concluding Remarks}
In this paper, we have shown how an area filling Peano curve represents a possible particle trajectory in the unit phase cell of a discrete phase space and continuous time relativistic quantum mechanical system. This is one of the first uses of this fascinating mathematical structure as a physical construct. Both first quantized Planck oscillators, first explored in part \textbf{I} of this work \cite{DasRCI}, and second quantized Klein-Gordon excitations were explored with respect to the Peano curve formalism.  Furthermore, the state space evolution of our Peano circle was shown to be analogous to the world-sheet evolution of closed strings. The geometric framework of this evolution was interpreted in terms of a product fibre bundle structure. Finally, extensions of our model to higher dimensions and striking similarities to popular dimensions used in traditional string theory were explored.   

\section*{Acknowledgements}
A.D. thanks Dr. Jack Gegenberg for some informal discussions.

\end{document}